# n-vicinity method and 1D Ising Model

## B.V. Kryzhanovsky, L.B. Litinskii

Scientific Research Institute for System Analysis RAS,
Center of Optical Neural Technologies
Nakhimov ave, 36-1, Moscow, 117218, Russia
kryzhanov@mail.ru, litin@mail.ru.

For a 1D Ising model, we obtained an exact expression for the spectral density in an n-vicinity of the ground state and explained why our n-vicinity method with the Gaussian approximation of the spectral density did not applicable in this case. We also found an analytical expression for the distribution of magnetization at an arbitrary temperature. When the temperature tends to zero the distribution of magnetization gradually flattens.

**Key words:** *spectral density, 1D Ising model, n-vicinity method, distribution of magnetization*

### 1. Introduction

The n-vicinity method may become a universal technique for calculation of the free energy of a thermodynamic system. The main idea of this method is as follows. The free energy can be easily calculated if we know the spectral density of the system. Since usually an exact expression for the spectral density of the system is unknown, we can try to replace it by a normal distribution with known values of its mean and variance. For Ising models on hypercubic high-dimensional lattices $d \geq 3$, the n-vicinity method allows us to obtain an analytical expression for critical temperature that describes correctly the results of computer simulations [1]. The calculated values of the critical exponents are also in good agreement with computer experiments [2].

In the same time for Ising models on hypercubic low-dimensional lattices, the n-vicinity method either does not applicable ($d=1$) or it predicts incorrectly the type of the phase transition ($d=2$). It will be useful to refine the n-vicinity method and extend the range of its applicability.

In the present paper, we analyze the 1D Ising model ($d=1$) for which we succeeded in obtaining an exact combinatorial expression for the spectral density of the states from n-vicinities. In Section 2, we obtain an analytical expression for the degeneracy of each value of the energy. This allows us to compare directly the true spectral density and its approximation by the normal distribution. We show that the normal distribution approximates correctly the central part of the spectral density near its maximum. However, it is not so good near the boundaries of the energy interval. In Section 3, we explain that this is the reason why the present version of the n-vicinity method fails to describe the properties of the 1D Ising system. In addition, in Section 4 we obtain the distribution of magnetization at an arbitrary value of the inverse temperature $\beta$. We show that when $\beta$ increases the distribution of the magnetization flattens or, in other word, tends to an equiprobable distribution. We use some physical reasoning to interpret this unexpected result. Conclusions and discussion are in Section 5. Technical details are in Appendixes.

### 2. Main formulas and exact spectral density

**1.** The main idea of the n-vicinity method is as follows [1, 4, and 5]. Let us fix an initial configuration $\mathbf{s}_0 \in \mathbf{R}^N$. We distribute other configurations over its n-vicinities $\Omega_n$ in such a way that in each $\Omega_n$ there are configurations that differ from $\mathbf{s}_0$ by opposite values of $n$ spins:

$$\Omega_n = \{\mathbf{s} : (\mathbf{s},\mathbf{s}_0) = N - 2n\}, \quad 0 \leq n \leq N. \quad |\Omega_n| = \binom{N}{n}.$$

We denote by $D_n(E)$ a true energy distribution of states from $\Omega_n$. As a rule, we do not know the function $D_n(E)$ but we can obtain exact expressions for the mean $E_n$ and the variance $\sigma_n^2$ of this unknown distribution [5]. In the framework of the n-vicinity method, we replace the unknown distribution $D_n(E)$ by a normal distribution with the mean $E_n$ and the variance $\sigma_n^2$:

$$D_n(E) \approx G_n(E) \sim \exp\left(-\frac{1}{2}\left(\frac{E-E_n}{\sigma_n}\right)^2\right).$$

The justification of this method we published in [5] and the boundaries of its applicability for Ising models were examined in [1].

**2.** In the case of a one-dimensional chain of spins with cyclic boundary conditions, the energy of the state **s** has the form [3]:

$$E(\mathbf{s}, H) = -\frac{(\mathbf{Js},\mathbf{s})}{2N} - H\frac{\sum_{i=1}^{N} s_i}{N} = E(\mathbf{s}) - H\cdot\left(1 - 2\frac{n}{N}\right), \quad (1)$$

where $H$ is an external magnetic field and $\mathbf{J}$ is a connection matrix of the 1D Ising model (see eq.(A1)). We choose the ground state of the spin system $\mathbf{s}_0 = (1,1,...1) \in R^N$ as an initial configuration. Then $E_0 = E(\mathbf{s}_0) = -1$.

In Appendix 1 we show (eqs.(A5) and (A6)) that for the states from the n-vicinity the energy $E(\mathbf{s})$ takes on $n$ different values $E(k)$ with the degeneracies $D(n,k)$, and

$$E(k) = -\left(1 - 4\frac{k}{N}\right), \quad D(n,k) = \frac{N\cdot k}{(N-n)n} C_{N-n}^{k} C_n^{k}, \quad k = 1, 2, ..., n. \quad (2)$$

The number of the vicinities $n$ runs from 1 to $N/2$.

The expressions (2) define completely the spectral density of the states from the n-vicinity. In what follows we tend $N$ to infinity and use a continuous analogue of these expressions. We introduce two numerical parameters $x = n/N$ and $y = k/N$ and rewrite eqs.(2) in a continuous form:

$$E(y) = -(1-4y), \quad D(x,y) = \frac{\exp\left[-N\cdot\left(xS\left(\frac{y}{x}\right) + (1-x)S\left(\frac{y}{1-x}\right)\right)\right]}{2\pi N\sqrt{x(1-x)(x-y)(1-x-y)}}, \quad x = \frac{n}{N} \in [0,1/2], \quad y = \frac{k}{N} \in [0,x], \quad (3)$$

where $S(a) = a\ln a + (1-a)\ln(1-a)$ is the Shannon function. The parameter $y$ defines the value of the energy and it varies inside the interval $[0, x]$. The parameter $x$ defines the value of the magnetization $m = 1 - 2x$ and it changes inside the interval $[0, 1/2]$.

We also need asymptotic expressions for the energies $E_x$ and the variances $\sigma_x^2$ averaged over n-vicinity[1]. In ref. [1], we showed that

$$E_x = \lim_{N\to\infty} E_n = -(1-2x)^2 \quad \text{and} \quad \sigma_x^2 = \lim_{N\to\infty} \sigma_n^2 = 16x^2(1-x)^2/N.$$

Then an asymptotic expression for the Gaussian approximation of the true density $D(x, y)$ has the form:

$$G(x,y) = \frac{e^{-\frac{1}{2}\left(\frac{E-E_x}{\sigma_x}\right)^2}}{\sqrt{2\pi\sigma_x}} = \frac{\exp\left[-N\left(S(x) + \frac{1}{2}\left(1 - \frac{y}{x(1-x)}\right)^2\right)\right]}{2\pi N\left(x(1-x)\right)^{3/2}}. \quad (4)$$

---

[1]In fact, the name x-vicinity is more correct here but we use the initial definition "n-vicinity".

We choose normalizations of the functions $D(x,y)$ and $G(x,y)$ in such a way that $\iint D(x,y)dxdy = \iint G(x,y)dxdy = 2^N$.

## 3. Comparison of true distribution with Gaussian approximation

Two notes have to be done. First, when comparing the true distribution of the energies and its Gaussian approximation, it is more convenient to use not the energy $E$ but the variable $y$. According to eq.(3), we have $y = (1+E)/4$. In what follow we compare the distributions $D(x,y)$ and $G(x,y)$ for different values of the parameter $x$. Second, we will compare not the densities $D(x,y)$ and $G(x,y)$ but their spectral functions defined as

$$\begin{aligned} d(x,y) &= \lim_{N\to\infty} \frac{\ln D(x,y)}{N} = -xS\left(\frac{y}{x}\right) - (1-x)S\left(\frac{y}{1-x}\right) \\ g(x,y) &= \lim_{N\to\infty} \frac{\ln G(x,y)}{N} = -S(x) - \frac{1}{2}\left(1 - \frac{y}{x(1-x)}\right)^2 \end{aligned}, \quad x \in [0, 1/2],\ y \in [0, x]. \tag{5}$$

The pre-exponential factors are not significant.

1. The graphs of the functions $d(x,y)$ for different values of the parameter $x$ are shown in the upper panel of Fig.1. Each curve reaches its maximum at the point $y_M = x(1-x)$ and then it decreases monotonically up to the boundary point $y = x$, where $d(x,x) = -(1-x)S\left(\frac{x}{1-x}\right)$. When the value of the parameter $x = 0.5$, the function $d(0.5, y) = -S(y)$. Here $S(y)$ is the Shannon function symmetric with respect to its maximum point $y_M = 0.25$.

One would expect that any other function $d(x,y)$ will also be symmetric about its maximum point $y_M$. In other words, the difference of the values of the function $d(x,y)$ at the points symmetric about $y_M$ will be identically zero:

$$\delta(x,y) = d(x,y) - d(x, 2y_M - y) \stackrel{?}{\equiv} 0, \quad \forall y \in [y_M, x]. \tag{6}$$

However, it is not the case. In the lower panel of Fig. 1, we show the differences $\delta(x,y)$ for the same values of the parameter $x$. They all are asymmetrical except the case when $x = 0.5$.

Finally let us discuss the behavior of the functions $d(x,y)$ near the boundaries of the interval $y \in [0, x]$. It is easy to obtain the following equalities:

$$\begin{aligned} &\lim_{y\to 0} d(x,y) = 0, & &\lim_{y\to x} d(x,y) = -(1-x)\cdot S\left(\frac{x}{1-x}\right), \\ &\lim_{y\to 0} d'_y(x,y) = \infty, & &\lim_{y\to x} d'_y(x,y) = -\infty, \\ &\lim_{y\to 0} d''_{yy}(x,y) = -\infty, & &\lim_{y\to x} d''_{yy}(x,y) = -\infty. \end{aligned} \tag{7}$$

At the left end of the interval, where $y = 0$, the first derivative of each function $d(x,y)$ tends to $+\infty$ and its second derivative tends to $-\infty$. At the right end of the interval, although the functions $d(x,y)$ themselves depend on $x$ their derivatives are independent of $x$. Such behavior of the functions $d(x,y)$ differs significantly from the behavior of the approximating function $g(x,y)$, which we will discuss in what follows.

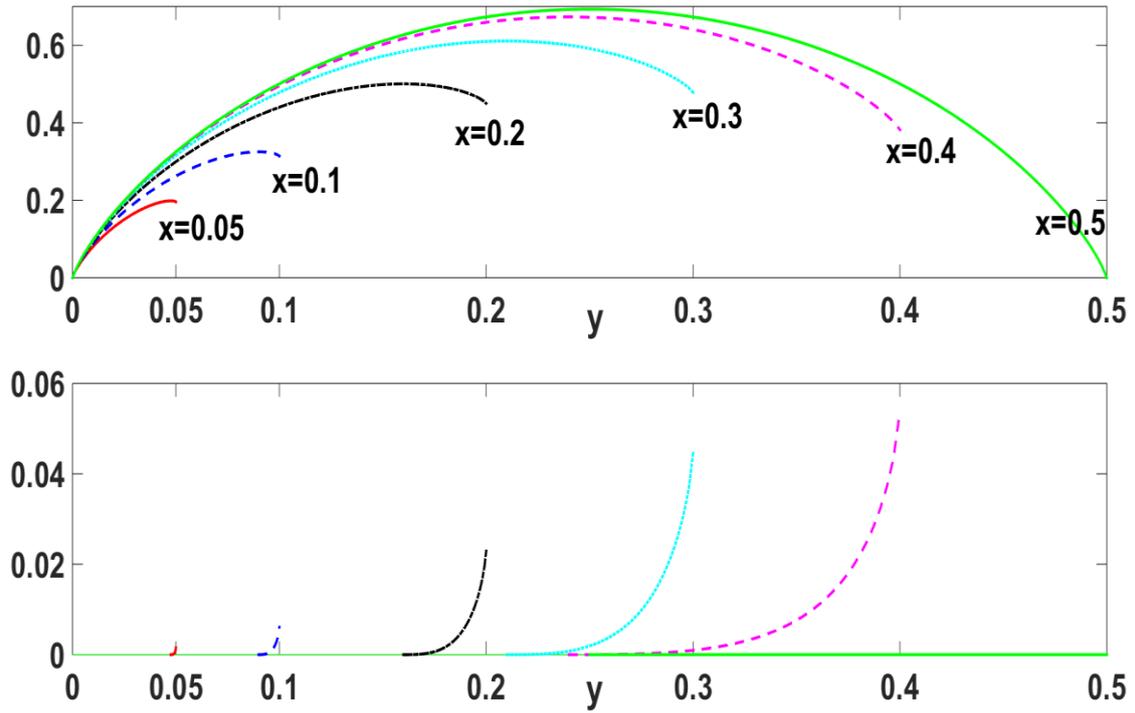

Fig.1. Upper panel: functions $d(x, y)$ (see eq.(5)) for different values of $x$;
lower panel : graphs of the differences $\delta(x, y)$ (see eq.(6)).

**2.** In Fig. 2 we present the graphs of the functions $d(x, y)$ and $g(x, y)$ for six values of the parameter $x$. We see (this is also clearly seen from eqs.(5)) that for any value of $x$ each function $g(x, y)$ has its maximum at the same point $y_M = x(1-x)$ as the function $d(x, y)$. The values of these functions at $y = y_m$ are also the same: $g(x, y_M) = d(x, y_M) = -S(x)$. This is a consequence of our approximation method of the spectral density in the n-vicinity: we approximate the true distribution in the neighborhood of its maximum by the Gaussian bell. That is near the maximum point $y_M$ the Gaussian curve approximate the true curve $d(x, y)$ rather accurately.

From the general reasoning, it is evident that $g(x, y)$ is a function symmetric about the maximum point $y_M$. In other word, we have the identity

$$g(x, y_M + \eta) - g(x, y_M - \eta) \equiv 0 \quad \forall \eta \in [0, x^2].$$

This is one of the differences of the functions $g(x, y)$ and $d(x, y)$ (compare with eq.(6)).

The behavior of the functions $g(x, y)$ and $d(x, y)$ also differs noticeably at the ends of the variation interval of $y$, that is near the points $y = 0$ and $y = x$. We can easily see this either from the graphs in Fig. 2 or from the formulas defining the boundary values of the function $g(x, y)$ (compare with eqs.(7)):

$$\lim_{y \to 0} g(x, y) = -S(x) - 0.5, \qquad \lim_{y \to x} g(x, y) = -S(x) - 0.5 \cdot \left(\frac{x}{1-x}\right)^2,$$

$$\lim_{y \to 0} g'_y(x, y) = \frac{1}{x(1-x)} > 0, \qquad \lim_{y \to x} g'_y(x, y) = -\frac{1}{(1-x)^2} < 0,$$

$$\lim_{y \to 0} g''_{yy}(x, y) = -\frac{1}{(x(1-x))^2} < 0, \qquad \lim_{y \to x} g''_{yy}(x, y) = -\frac{1}{(x(1-x))^2} < 0.$$

At the boundaries of the interval, the function $g(x, y)$ and its derivatives with respect to $y$ depend on $x$. When $x \neq 0$, the derivatives do not tend to infinity (compare with eqs. (7)). Let us explain why the behavior of the functions $g(x, y)$ and $d(x, y)$ near the ends of the interval is so important.

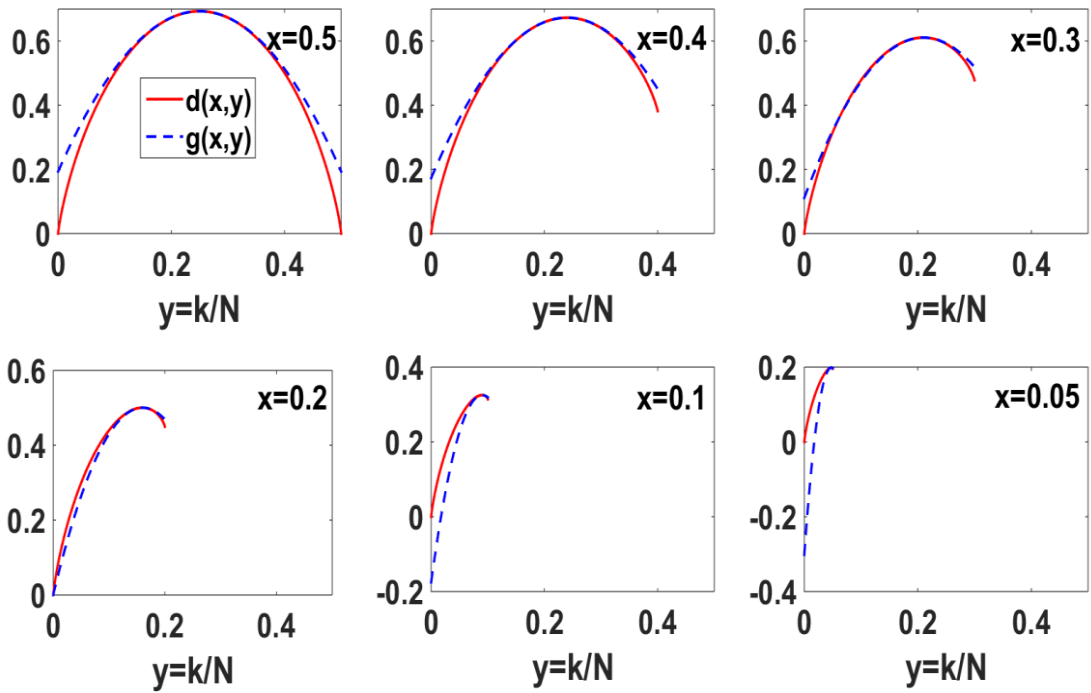

Fig. 2. Graphs of functions $d(x, y)$ (solid lines) and their Gaussian approximations $g(x, y)$ (dashed lines; see eq.(5)) for $x = 0.5, 0.4, 0.3, 0.2, 0.1, 0.05$.

**3.** The function $D(x, y)$ (see eq.(3)) describes the distribution of energies for the part of the states only, namely for the states that belong to the n-vicinity. That is the reason why below we refer to $D(x, y)$ as *a partial spectral density* or *a partial density*. We summarize the densities $D(x, y)$ over all the n-vicinities and obtain *a total spectral density* (in short *a total density*) $D(y) = \int D(x, y)dx$, which describes the distribution of energies for all set of $2^N$ configurations. Let us write it as an exponent: $D(y) = \exp(N \cdot d(y))$. In the same way we summarize the Gaussian approximations $G(x, y)$ and introduce the Gaussian approximation of the total density: $G(y) = \int G(x, y)dx = \exp(N \cdot g(y))$. With the aid of the functions $D(y)$ and $G(y)$ we explain why for the 1D Ising system the normal distribution $G(x, y)$ cannot be used when approximating the partial density.

Let us describe the main points of the calculation of the total density (for details see Appendix 2). When integrating $D(x, y)$ over $x$, we obtain

$$D(y) = \int_y^{1-y} D(x, y)dx = \int \exp(N \cdot d(x, y))dx \sim \exp\left(N \cdot \max_x d(x, y)\right) = \exp(N \cdot d(0.5, y)). \qquad (8)$$

The last equation in the chain (8) follows from the fact that for any $x$ the maximum point of $d(x, y)$ is $x = 0.5$ (see Appendix 2). In other words, we have

$$d(y) = d(0.5, y) = -S(2y). \qquad (9)$$

In the same way we show that $G(y) = \int_y^{1-y} G(x, y)dx = ... = \exp(N \cdot g(0.5, y))$ and

$$g(y) = g(0.5, y) = \ln 2 - \frac{(1-4y)^2}{2}. \qquad (10)$$

In Ref. [6] we showed that if $E_0$ is the energy of the ground state and the total density is an exponent $\exp(N \cdot \Psi(E))$, where $E \in [E_0, |E_0|]$, the spectral function $\Psi(E)$ has to fulfill the following conditions:

- $\lim_{E \to E_0} \Psi(E) = 0$. If not, we obtain a multidegenerate ground state and this is impossible in the Ising models.

- $\lim_{E \to E_0} \Psi'(E) = \infty$. If not, the system reaches the ground state at a finite temperature and this contradicts to the general physical principles.

In the case of the 1D Ising system we have $E_0 = -1$, and $y = (1+E)/4 \to 0$ when $E \to E_0$. In the terms of the parameter $y$ the above-said means that a function $\psi(y)$ can play the role of the spectral function of the total density if the conditions $\lim_{y \to 0} \psi(y) = 0$ and $\lim_{y \to 0} \psi'(y) = \infty$ are fulfilled. The function $d(y)$ (9) meets these conditions. This statement can be easily confirmed by means of direct calculations. On the contrary, the approximating function $g(y)$ (10) does not satisfy any of them. And this is also easy to see. This means in the case of the 1D Ising system, we cannot use the Gaussian density to approximate the true partial density in the n-vicinity and consequently the n-vicinity method is inapplicable.

### 4. Partition function and distribution of magnetization

**1.** With the aid of the true spectral density $D(x, y)$ (see eq.(3)) we can calculate the partition function $Z(\beta, H)$:

$$Z_N(\beta, H) = \sum_{n=0}^{N} \sum_{s \in \Omega_n} D(n, k) e^{-\beta N E(s, H)} \to \frac{N}{2\pi} 2 \int_0^{1/2} \frac{dx}{\sqrt{x(1-x)}} \int_0^x \frac{dy \cdot \exp(-N \cdot \varphi(x, y, \beta, H))}{\sqrt{(x-y)(1-x-y)}} \sim \exp(-N \cdot \varphi(x_c, y_c, \beta, H)), \qquad (11)$$

where $\varphi(x, y, \beta, H)$ has the form (see eqs.(1-3))

$$\varphi(x, y, \beta, H) = x \cdot S\left(\frac{y}{x}\right) + (1-x) \cdot S\left(\frac{y}{1-x}\right) + \beta \cdot (4y - 1 + H \cdot (2x - 1)),$$

and the values of $y_c$ and $x_c$ correspond to the global minimum of $\varphi(x, y)$. (For simplicity in what follows we do not indicate the dependence of different functions on the parameters $\beta$ and $H$. For example, we will write $\varphi(x, y)$ in place of $\varphi(x, y, \beta, H)$ and so on.)

In equation (11) $y_c$ is the solution of equation $\partial \varphi(x, y)/\partial y = 0$. It is easy to see that

$$y_c(x) = \frac{2x(1-x)}{1+q(x)} = \frac{q(x) - 1}{2b}, \text{ where } q(x) \equiv q(x, \beta) = \sqrt{1 + 4x(1-x) \cdot b} \text{ and } b = e^{4\beta} - 1. \qquad (12)$$

Let us note that the coordinate of the saddle point $y_c(x)$ does not depend on the value of the magnetic field.

To find $x_c$ it is necessary to solve the equation $\partial \varphi(x, y_c(x))/\partial x = 0$. After tedious algebra, we can rewrite it as

$$\varphi'_x(x, y_c(x)) = \ln\left(1 - \frac{y_c(x)}{x}\right) - \ln\left(1 - \frac{y_c(x)}{1-x}\right) + 2\beta H = 0. \tag{13}$$

Suppose that the magnetic field $H = 0$. Then the solution of equation (13) is $x_c \equiv x_0 = 1/2$, and we obtain

$$y_c(x_c) = \frac{1}{2(1+e^{2\beta})}, \quad \varphi(x_c, y_c) = S\left(\frac{1}{1+e^{2\beta}}\right) - \beta\frac{e^{2\beta}-1}{e^{2\beta}+1}. \tag{14}$$

Now the expression for the free energy is

$$f(\beta) = -\frac{1}{\beta}\lim_{N\to\infty}\frac{\ln Z_N}{N} = \frac{\varphi(x_c, y_c)}{\beta} = \frac{1}{\beta}S\left(\frac{1}{1+e^{2\beta}}\right) - \frac{e^{2\beta}-1}{e^{2\beta}+1} \equiv -\frac{1}{\beta}\left(1 + \ln(1+e^{-2\beta})\right).$$

In this chain of equalities, the last identity corresponds to the form of the partition function commonly used in literature [3].

**2.** The magnetization is an important characteristic of the system. It is simply related to the variable $x$: $m = 1 - 2x$. If in eq.(11) we restrict ourselves by integration over $y$ only, we obtain a function

$$P_N(x,\beta) = \sqrt{\frac{N}{2\pi}}\int_0^x \frac{\exp(-N\cdot\varphi(x,y))\cdot dy}{\sqrt{(x-y)(1-x-y)}} = \frac{\exp(-N\cdot\varphi(x,y_c))}{\sqrt{2\pi N x(1-x)}\sqrt{q(x,\beta)}},$$

which is proportional to the number of states with the magnetization $m = 1 - 2x$ at a given inverse temperature $\beta$. Dividing this function by the partition function $Z(\beta)$ we obtain the density of distribution of the magnetization $m$ at a given $\beta$:

$$p_N(x,\beta) = \frac{P_N(x,\beta)}{Z_N(\beta)} = \frac{\exp\left(-N\left[\varphi(x,y_c(x)) + \beta + \ln\left(1+e^{-2\beta}\right)\right]\right)}{\sqrt{2\pi N x(1-x)}\sqrt{q(x,\beta)}}, \quad x \in [0,1/2], \; \beta \in [0,\infty). \tag{15}$$

Generally speaking, the value of $x$ cannot be infinitely small. In Appendix 2, we show that since in our calculations we use the Stirling formula, the value of $n = xN$ has to be sufficiently large. Let us suppose that $n_{\min} \sim 10^2 - 10^3$. Then $x_{\min}$ is sufficiently small but does not equal to zero. This means that for small $x$ there is no singularity of the function $p_N(x,\beta)$ and we have

$$\lim_{x\to 0} p_N(x,\beta) \sim \frac{\exp(-N(1+e^{-2\beta}))}{\sqrt{2\pi n_{\min}}}. \tag{16}$$

This equation will be useful in what follows.

In Fig. 3 we show the graphs of the density of the states $p_N(x,\beta)$ defined by eq.(15) and the corresponding arguments of the exponent $\ln(p_N(x,\beta))/N$ (right panel) for different values of $\beta$. In this figure the number of spins is $N = 1000$. When $\beta$ is about zero, $\beta \sim 0 \Rightarrow T = 1/\beta \sim \infty$, the magnetization of almost all the states is inside a narrow region near zero. We see a peak near the value $x_0 = 0.5$, where $m = 0$. When $\beta$ increases (that is when the temperature $T$ decreases) the height of the peak decreases, it becomes wider and finally flattens. This behavior of the distribution of magnetization is not obvious.

Indeed, when the temperature of the system decreases the system tends to the ground state $\mathbf{s}_0$ whose magnetization $m = 1$ and consequently $x$ has to be equal to zero. It would seem that when $\beta$ increases at the left boundary of the interval (near $x = 0$) a peak of the density distribution has to be formed. However, there is not the case. In the right panel, we see that when $\beta$ increases the curve $\ln(p_N(x,\beta))/N$ flattens and its left end raises up but all the same, it remains lower than the right end.

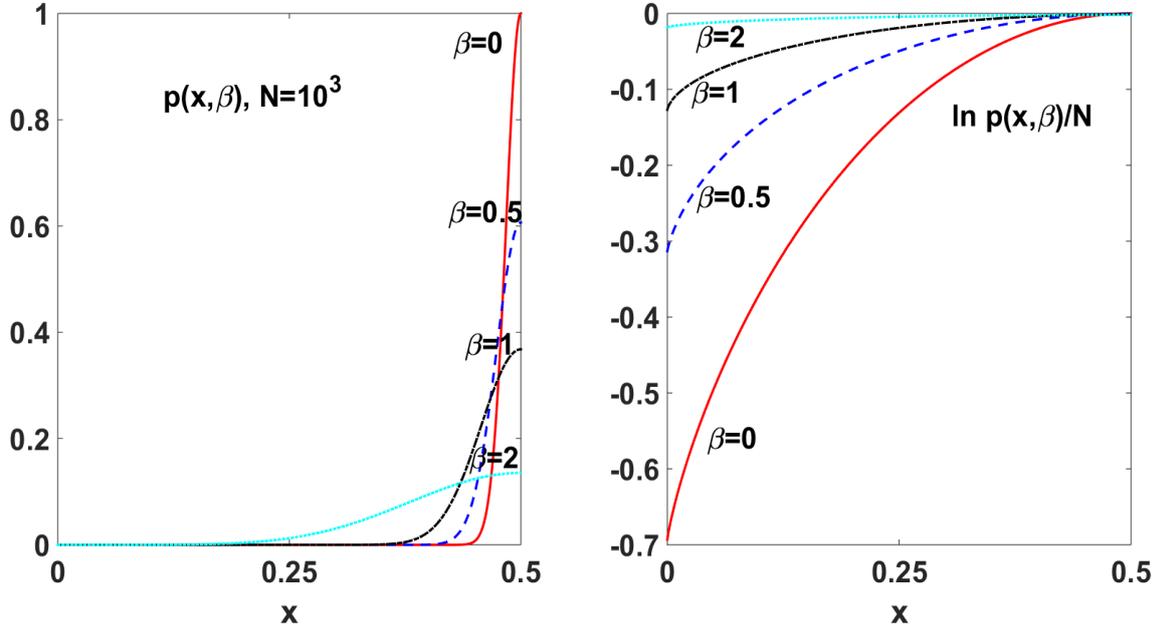

Fig. 3. Density distribution of magnetization $p_N(x,\beta)$ (left panel) and $\ln(p_N(x,\beta))/N$ (right panel) for $\beta = 0, 0.5, 1, 2$.

To show that no peak should be at the left end of the interval, let us estimate the value of $p_N(x,\beta)$ at the right end of the interval ($x \to 1/2$). With regard to eq.(14) we see that in eq.(15) the value of the argument of the exponent tends to zero and the value of $q(x)$ tends to $e^{2\beta}$ (see eq.(12)). Consequently, we have

$$\lim_{x \to 1/2} p_N(x,\beta) = \sqrt{\frac{2}{\pi N}} e^{-\beta}. \tag{17}$$

Comparing this expression with eq.(16) we see that when $N \to \infty$ at the left end of the interval the argument of the exponent is $-N(1+e^{-2\beta})$ for each value of $\beta$. This value is much less than the value of the argument of the exponent at the right end of the interval where it is equal to $-\beta$. The left end of the curve $p_N(x,\beta)$ always lower than its right end. Consequently, when $\beta \to \infty$ the distribution of the magnetization is uniform.

Let us explain this result. When $\beta \to \infty$, the system tends to the ground state $\mathbf{s}_0$ and its energy is close to $E_0 = -1$. In other words, the values of energy are localized near $E_0$. However, it does not mean that the values of the magnetization have to group near one. Indeed, if *only* the states from a small vicinity near $\mathbf{s}_0$ have the energies close to $E_0$, the values of magnetization have to group near one when $\beta \to \infty$. However, it is easy to see that in any n-vicinity of $\mathbf{s}_0$ there are $N$ configurations whose energies equal to $E(1) = -(1-4/N)$. They differ from $E_0$ by an infinitely small value. To confirm this statement in the expression (2) for the number of states $D(n,k)$ we substitute $k=1$ that corresponds to the energy $E(1)$. We obtain that $D(n,1) = N$. Thus, we see that configurations whose energies are practically equal to $E_0$ are not localized near the ground state $\mathbf{s}_0$, but distributed over all the configuration space. This explains why the curves $p_N(x,\beta)$ flatten when $\beta$ increases and we obtain a nearly uniform distribution.

## 5. Discussion and conclusions

The spectral density defines the partition function as well as many other important macroscopic characteristics of the system. However, the problem of calculation of the partition function is very difficult and it is solved only in a few specific cases. We succeeded in deriving the exact combinatorial expression for the spectral density in the n-vicinity of the ground state. This allowed us to compare the exact spectral density with its approximation by the Gaussian density that is usually used in the n-vicinity method. We found that the Gaussian density approximates well the central part of the exact distribution but the behavior of these functions differs notable at the ends of the interval. This is the reason why the n-vicinity method in its present "Gaussian" form does not work in the case of the 1D Ising model.

In the future, we have to explain why the Gaussian approximation allows us to obtain correct results for the high-dimensional lattices. When $d \geq 3$, using the n-vicinity method we obtained a simple analytical expression for the critical temperature that reasonable fits to generally recognized results of computer simulations [1,2]: It is difficult to consider this coincidence as accidental.

We also plan to examine another question: is it possible to adapt the n-vicinity method to examination of low-dimensional lattices ($d < 3$)? The most interesting is the 2D Ising model in the external magnetic field. This problem that was formulated more than half a century ago is not solved yet.

In addition, even in the case of the 1D Ising model, which was examined in detail, the obtained exact expression for the spectral density allowed us to obtain a new result: when the temperature tends to zero the distribution of magnetization tends to be uniform. We hope to generalize the exact expression for the spectral density obtained for the 1D Ising model to the case of the 2D Ising model.

The work was supported by the RBRF grant 18-07-00750.

## APPENDIX 1

**1.** Let us write the connection matrix $\mathbf{J}$ of the 1D Ising model as a sum of a one-step shifting matrix $\mathbf{T}$ and the transposed matrix $\mathbf{T}^+$:

$$\mathbf{J} = \begin{pmatrix} 0 & 1 & 0 & 1 \\ 1 & 0 & \ddots & 0 \\ 0 & \ddots & \ddots & 1 \\ 1 & 0 & 1 & 0 \end{pmatrix} = \mathbf{T} + \mathbf{T}^+, \text{ где } \mathbf{T} = \begin{pmatrix} 0 & 0 & 0 & 1 \\ 1 & 0 & \ddots & 0 \\ 0 & \ddots & \ddots & 0 \\ 0 & 0 & 1 & 0 \end{pmatrix}. \tag{A1}$$

The interaction constant is equal to one: $J_{ii+1} = 1$.

Let $\mathbf{e}_i$ be the $i$-th unit vector in the space $\mathbf{R}^N$: $(\mathbf{e}_i, \mathbf{e}_j) = \delta_{ij}$, where $\delta_{ij}$ is the Kronecker symbol and $i = 1,...,N$. The matrix $\mathbf{T}$ transforms the $i$-th unit vector in the $(i+1)$-th unit vector: $\mathbf{T}\mathbf{e}_i = \mathbf{e}_{i+1}$. In addition we suppose that $\mathbf{T}\mathbf{e}_N = \mathbf{e}_1$. By $F(\mathbf{s})$ we denote the value of the quadratic form: $F(\mathbf{s}) = (\mathbf{T}\mathbf{s}, \mathbf{s})$. Then the energy of the state $\mathbf{s}$ is equal to:

$$E(\mathbf{s}) = -\frac{(\mathbf{J}\mathbf{s},\mathbf{s})}{2N} = -\frac{F(\mathbf{s})}{N}. \tag{A2}$$

In what follows it is more convenient to use not the energy $E(\mathbf{s})$ but the function $F(\mathbf{s})$.

We choose the ground state $\mathbf{s}_0 = (1,1,...,1) = \sum_{i=1}^{N} \mathbf{e}_i$ as an initial configuration and by $\mathbf{s}_{i_1 i_2 .. i_n}$ we denote the configurations from the n-vicinity of $\mathbf{s}_0$ supposing that spins «-1» are at the positions $i_1, i_2,...,$ and $i_n$:

$$\mathbf{s}_{i_1 i_2 .. i_n} = \mathbf{s}_0 - 2(\mathbf{e}_{i_1} + \mathbf{e}_{i_2} ... + \mathbf{e}_{i_n}).$$

It is easy to see that

$$F(\mathbf{s}_{i_1i_2..i_n}) = N - 4 \cdot n + 4 \cdot \left(\mathbf{e}_{i_1+1} + \mathbf{e}_{i_2+1} + ... + \mathbf{e}_{i_n+1}, \mathbf{e}_{i_1} + \mathbf{e}_{i_2} + ... + \mathbf{e}_{i_n}\right). \quad (A3)$$

By

$$\Delta_{i_1i_2..i_n} = \left(\mathbf{e}_{i_1+1} + \mathbf{e}_{i_2+1} + ... + \mathbf{e}_{i_n+1}, \mathbf{e}_{i_1} + \mathbf{e}_{i_2} + ... + \mathbf{e}_{i_n}\right) \quad (A4)$$

we denote the scalar product in eq.(A3).

The set of different $\Delta_{i_1i_2..i_n}$ defines the energy spectrum in the n-vicinity. It is not difficult to understand the construction of these scalar products. At first, suppose that the indices $i_j$ are in series: $i_2 = i_1 + 1, i_3 = i_2 + 1, ...$, and $i_n = i_{n-1} + 1$. It is easy to see that in this case the value of $\Delta_{i_1i_2..i_n}$ is equal to $n-1$. Next, let the set of indices falls into two isolated subgroups. Inside each of the group the indices are in series so that the first group is $i, i+1, .., i+a$ and the second group is $k, k+1, .., k+b$. Evidently, $a+b+2 = n$ but there is a gap between the value of the index $i+a$ and the index $k$: $k > i+a+1$. It can easily be checked that independent of the structures of the groups the value of $\Delta_{i_1i_2..i_n}$ is equal to $n-2$. Similarly, when the set of the indices $i_1, i_2, ..., i_n$ falls into three isolated subgroups inside which the indices are in series the value of the scalar product $\Delta_{i_1i_2..i_n}$ is equal to $n-3$ and it is independent of the structures of the groups, and so on. When the set of indices $i_1, i_2, ..., i_n$ falls into $k$ isolated subgroups (inside each of which the indices are in series) the value of $\Delta_{i_1i_2..i_n}$ is equal to $n-k$. When $n \leq N/2$ the maximal possible number of such isolated subgroups is equal to $k_{max} = n$. Using equations (A2), (A3) and (A4), we obtain that the energies of the states from the n-vicinity take exactly $n$ different values:

$$E(k) = -(1 - 4k/N), k = 1, 2, ... n, \text{ where } n = 1, 2, .., N/2. \quad (A5)$$

**2.** Now let us find the degeneracies of the energy levels $E(k)$ in the n-vicinity. By $D(n,k)$ we denote the number of the state in the n-vicinity whose energy is equal to $E(k)$.

The 0-vicinity consists of the configuration $\mathbf{s}_0$ itself, $F(\mathbf{s}_0) = (\mathbf{s}_0, \mathbf{s}_0) = N$ and we can write: $E_0 = -1$ and $D(0,0) = 1$.

The 1-vicinity of $\mathbf{s}_0$ consists of $N$ configurations that differ from $\mathbf{s}_0$ by different value of one spin only. Consequently, for all $N$ configurations from the 1-vicinity the energies are equal to the same value $E(1)$ and $D(1,1) = N$.

It is easy to see that $C_N^2$ configurations from the 2-vicinity of $\mathbf{s}_0$ falls into two subgroups. First, there are $N$ configurations where two «-1» spins are in series. Let $E(1)$ be the energy of the states from this subgroup. Second, all other configurations belong to the second subgroup. The energy of these states is equal to $E(2)$. Consequently,

$$D(2,1) = N, \; D(2,2) = \frac{N(N-3)}{2}.$$

It is not difficult to show that in the 3-vicinity there are exactly $N$ configurations that are characterized by the energy $E(1)$, $N(N-4)$ configurations that are characterized by the energy $E(2)$, and the energy of all other configurations is equal to $E(3)$. Then we obtain:

$$D(3,1) = N, \; D(3,2) = N(N-4), \; D(3,3) = \frac{N(N-4)(N-5)}{3!}.$$

From now on, the calculations are more cumbersome but with the aid of a straightforward analysis for the 4-vicinity and 5-vicinity we obtain the following results:

$$D(4,1) = N, \quad D(4,2) = \frac{3}{2}N(N-5), \quad D(4,3) = \frac{N(N-5)(N-6)}{2}, \text{ and } D(4,4) = \frac{N(N-5)(N-6)(N-7)}{4!};$$

$$D(5,1) = N, \quad D(5,2) = 2N(N-6), \quad D(5,3) = N(N-6)(N-7), \quad D(5,4) = \frac{N(N-6)(N-7)(N-8)}{3!}, \text{ and }$$

$$D(5,5) = \frac{N(N-6)(N-7)(N-8)(N-9)}{5!} \text{ respectively.}$$

An analyses of the expressions for the values of $D(n,k)$ obtained for the five vicinities of the ground state allowed us to write a combinatorial formula that we confirmed by computer simulations for different values of $N, n$ and $k$:

$$D(n,k) = NC_{N-n-1}^{k-1} \cdot C_n^k / n = \frac{Nk}{(N-n)n} C_{N-n}^k C_n^k, \quad \begin{array}{l} n = 1, 2, ... N/2; \\ k = 1, 2, ... n. \end{array} \tag{A6}$$

**APPENDIX 2**

Let us derive the expression for the total spectral density $G(y)$ by integrating the partial spectral density $G(x, y)$ over $x$ (see eq.(4)):

$$G(x, y) = \frac{e^{-N \cdot \varphi(x,y)}}{2\pi N (x(1-x))^{3/2}}, \text{ where } \varphi(x, y) = S(x) + \frac{1}{2}\left(1 - \frac{y}{x(1-x)}\right)^2.$$

Here $S(x)$ is the Shannon function that appears due to application of the Stirling formula. Using the Laplase method, we obtain a following chain of equalities:

$$G(y) = N \cdot \int_y^{1-y} G(x, y) dx = \frac{2}{2\pi} \int_y^{1/2} \frac{e^{-N \cdot \varphi(x,y)} dx}{(x(1-x))^{3/2}} = \frac{(2\pi N)^{-1/2} \cdot e^{-N \cdot \varphi(x_{min}, y)}}{(x_{min}(1-x_{min}))^{3/2} \cdot \sqrt{\varphi''_{xx}(x_{min}, y)}}, \tag{A7}$$

where $x_{min}$ is the point of the global minimum of the function $\varphi(x, y)$ with respect to the variable $x$. Generally speaking, $x_{min}$ depends on the value of the variable $y$ but we will see that this is not the case here. Indeed, the equation for $x_{min}$ has the form

$$\varphi'_x(x, y) = \ln \frac{x}{1-x} + \frac{y(1-2x)}{(x(1-x))^2}\left(1 - \frac{y}{x(1-x)}\right) = 0. \tag{A8}$$

It is evident that the point $x_0 = 1/2$ is a solution of equation (A8) and for all $y$ the second derivative $\varphi''_{xx}(x_0, y) = 4(1 - 8y + 32y^2)$ is positive. Consequently, $x_0$ is always the minimum point of the function $\varphi(x, y)$. It can be shown that if $y > 0.053$, the function $\varphi(x, y)$ has only one minimum at the point $x_0$. When $y < 0.053$, the equation (A8) has two additional solutions. They are the minimum point $x_m$ and the maximum point $x_M$. At that, $0 < x_m < x_M < x_0$. The direct analysis shows that the minimum $\varphi(x_0, y)$ always deeper than the minimum $\varphi(x_m, y)$ at the left end of the interval. This means that for any $y$ the global minimum of the function $\varphi(x, y)$ with respect to the variable $x$ always is at the point $x_0 = 1/2$.

When substituting $x_0$ in place of $x_{min}$ in the expression (A7), we finally obtain

$$G(y) = \frac{4 \cdot \exp\left(N \cdot \left(\ln 2 - \frac{(1-4y)^2}{2}\right)\right)}{\sqrt{2\pi N} \cdot \sqrt{1 - 8y + 32y^2}}, \text{ where } y \in [0, 1/2]. \tag{A9}$$

The similar calculations show that the partial spectral density $D(x, y)$ has only one minimum at the point $x_0 = 1/2$ (see eq.(3)). Then we obtain

$$D(y) = \int_{y}^{1-y} D(x, y)dx = \frac{2 \cdot \exp(-N \cdot S(2y))}{\sqrt{2\pi N} \cdot \sqrt{2y(1-2y)}} = 2C_N^{2y \cdot N}. \qquad (A10)$$

The last expression coincides with the result that was obtained in the end of 1930s with the aid of other reasoning [7].